Analytic Criteria for Power Exhaust in Divertors due to Impurity Radiation

D. Post[1], N. Putvinskaya[1], F. W. Perkins[1] and W. Nevins[2]

[1]ITER Joint Central Team, San Diego, CA

[2]Lawrence Livermore National Laboratory

ABSTRACT

Present divertor concepts for next step experiments such ITER and TPX rely upon impurity and hydrogen radiation to transfer the energy from the edge plasma to the main chamber and divertor chamber walls[1]. The efficiency of these processes depends strongly on the heat flux, the impurity species, and the connection length. Using a database for impurity radiation rates constructed from the ADPAK code package[2, 3], we have developed criteria for the required impurity fraction, impurity species, connection length and electron temperature and density at the mid-plane. Consistent with previous work[4, 5], we find that the impurity radiation from coronal equilibrium rates is, in general, not adequate to exhaust the highest expected heating powers in present and future experiments. As suggested by others[6, 7], we examine the effects of enhancing the radiation rates with charge exchange recombination[8]and impurity recycling, and develop criteria for the minimum neutral fraction and impurity recycling rate that is required to exhaust a specified power. We also use this criteria to find the optimum impurity for divertor power exhaust.

I. INTRODUCTION

The next generation of tokamak experiments are expected to have sufficiently high peak loads that the divertors for these experiments are being designed to maximize the transfer of the heating power to the divertor side walls by atomic processes including impurity radiation. We have integrated the equation for heat conduction equation along the field lines in the scrape-off layer including impurity radiation losses with enhancements due to impurity recycling and charge exchange recombination to develop criteria for the impurity fraction, impurity species, parallel peak heat flux, and connection length required to radiate the energy and for the plasma to "detach".

II. MODEL:

Using electron pressure balance along the field line we can cast the second order equation for heat conduction along the field lines as two first order equations with respect to the electron temperature:

$$\frac{\partial Q_\|}{\partial x} = -n_e n_z L_Z(T_e) \; \vdots \; Q_\| = -\kappa_o T_e^{2.5} \frac{\partial T_e}{\partial x} \; \vdots \; p_e = n_e T_e \Rightarrow$$

$$Q_\| \frac{\partial Q_\|}{\partial x} = -n_e n_z L_Z Q_\| \approx \kappa_o n_e n_z L_Z T_e^{2.5} \frac{\partial T_e}{\partial x} \Rightarrow$$

$$\frac{\partial Q_\|^2}{\partial T_e} \approx \frac{p_e^2}{T_e^2} 2\kappa_o f_z L_Z T_e^{2.5} \approx p_e^2 2\kappa_o f_z L_Z T_e^{0.5} \quad (1)$$

$$\text{where } \kappa_o \approx \frac{3.1 \times 10^9}{Z_{eff} \ln \Lambda} \left( \frac{erg}{cm \; s \; eV^{3.5}} \right)$$

$$\frac{\partial Q_\|^2}{\partial T_e} = p_e^2 2\kappa_o f_z L_Z T_e^{0.5} \text{ and } \frac{\partial x}{\partial T_e} = \frac{-\kappa_o T_e^{2.5}}{Q_\|} \quad (2)$$

These two equations have a natural set of variables of the form:

$$\tilde{q} = \frac{Q_\|}{n_s} \sqrt{\frac{Z_{eff} \ln \Lambda}{f_Z}} \text{ and } \xi = n_s \, x \, \sqrt{Z_{eff} \ln \Lambda \, f_Z} \quad (3)$$

in which the equations have the form:

$$\frac{\partial \tilde{q}^2}{\partial T_e} = T_s^2 2\kappa_o \sqrt{T_e} L_Z(T_e) \quad \text{and} \quad \frac{\partial \xi}{\partial T_e} = \frac{\kappa_o T_e^{2.5}}{\tilde{q}} \quad (4)$$

These equations can be cast into a set of practical units by expressing $Q_\|$ in GW/m$^2$, $n_s$ in $10^{20}$ m$^{-3}$, $T_e$ in 100 eV, $f_z$ in %, L(T) in $10^{-25}$ watts-cm$^3$ and x in 100 m and obtain the two equations:

$$\frac{\partial \tilde{q}^2}{\partial \tau_e} = 0.52 \, \tau_s^2 \sqrt{\tau_e} \Omega_Z(T_e) \quad \text{and} \quad \frac{\partial \xi}{\partial \tau_e} = 0.26 \frac{\tau_e^{2.5}}{\tilde{q}} \quad (5)$$

where $\tau \equiv \frac{T_e}{100 eV} : Z_v \equiv \frac{Z_{eff} \ln \Lambda}{12}$. These can be integrated to yield:

$$\tilde{q}^2 = \tilde{q}_o^2 + 0.52 \, \tau_s^2 \int_0^{\tau_s} \sqrt{\tau_e} \Omega_Z(\tau_e) d\tau_e \quad \text{and}$$

$$\xi = 0.26 \int_{\tau_{min}}^{\tau_s} \frac{\tau_e^{2.5} d\tau_e}{\sqrt{\tilde{q}_o^2 + 0.52 \, \tau_e^2 \int_0^{\tau_e} \sqrt{\tau_t} \Omega_Z(\tau_t) d\tau_t}} \quad (6)$$

$\tilde{q}_o$ is the heat flux that reaches the divertor plates. $\tau_{min}$ is chosen to be ~ a few eV to avoid divergence of the integral for very low temperatures where $L(T_e)$ goes to zero very rapidly. The first equation in (5) determines whether or not the radiation losses can exceed the heat input. The second equation in (5) defines the distance along the field line to the point of detachment. The larger the integral over $\Omega T^{0.5}$, the shorter the distance is for detachment to occur.

The plasma detaches when all of $Q_{\parallel}$ can be radiated (Equation 7) by an impurity with a concentration of $f_z$, with a scrape-off density of $n_s$ and temperature of $T_s$.

$$Q_{\parallel} \leq n_s T_s \sqrt{2\hat{\kappa}_o \frac{f_z}{1+f_z Z(Z-1)} \int_0^{T_s} L_Z T_e^{0.5} dT_e}$$

$$\text{where } \hat{\kappa}_o \approx \frac{3.1 \times 10^9}{\ln \Lambda} \left( \frac{erg}{cm\ s\ eV^{3.5}} \right) \tag{7}$$

The limit on $Q_{\parallel}$ can be increased by:

- Increasing $n_s$

    However, $n_s$ is determined by the main plasma conditions. Usually $n_s \leq n_{central}$.

- Increase $f_z$

    $f_z$ is limited by the need to keep the impurity concentration below the point where radiation losses from the core plasma severely impact energy confinement[11]. In any case $f_z$ is limited to $1/Z$ so that $\sqrt{\frac{f_z}{1+f_z Z(Z-1)}}$ is limited to $1/Z$ for a 100% impurity plasma.

- Increase $L_z(T_e)$:

    This can be done with charge exchange recombination[8] and rapid impurity recycling and is the subject of this paper. The choice of the radiating element strongly affects $L_z(T_e)$ but the trade-off is constrained by the need to avoid excessive central radiation losses.

- Increase $T_s$:

    Equation 7 does not explicitly include any dependence on the distance along the field lines from the main plasma to the divertor plate. However, $T_s$ implicitly depends on this distance because $T_s$ is derived from the heat conduction equation (Equation 8) along the field lines:

$$-\kappa_o \frac{\partial T_e^{3.5}}{\partial x} = Q_{\parallel} \therefore T_s^{3.5} = \frac{1}{\kappa_o} \int_{divertor}^{separatrix} Q_{\parallel} dT_e \approx \frac{Q_{\parallel} L}{\kappa_o} \text{ so } T_s \approx \left( \frac{7 Q_{\parallel} L}{2 \kappa_o} \right)^{2/7} \tag{8}$$

For a fixed $Q_\|$, $T_s$ is proportional to $L^{2/7}$ so that $T_s$ can be increased by increasing the connection length. Thus detachment for almost any set of conditions can be obtained if L is made sufficiently large. However, practical limits on L set it to ~ $1.5\pi q_{\psi 95} R$.

- $Q_\|$ itself can be made smaller by increasing the radial decay width of the power, but is usually not an adjustable design parameter. Operation at higher $q_\psi$ will also increase the connection length and lower $Q_\|$.

Equation 6 defines the connection length $\xi$ in terms of $T_s$ and the other parameters. If the connection length from 6 is longer than the connection length in the experiment, then detachment does not occur. This is, however, not as useful as it might seem since one must know something about L to compute $T_s$. However, one can use both equations in (5) $\tilde{q} = \tilde{q}(\tau_S)$ and $\xi = \xi(\tau_S)$ to find $\tilde{q} = \tilde{q}(\tau_S(\xi)) = \tilde{q}(\xi)$ to obtain the maximum heat flux as a function of the field line length.

The values of the impurity emission function $L_Z(T_e)$ were computed using the ADPAK code which is based on an average ion model[2, 3]. Charge exchange recombination effects[2] were characterized in terms of the neutral fraction ($n_0/n_e$). Impurity recycling effects were parametrized in terms of $n_e\tau_{recy}$, the product of the recycling time and electron density. The radiation rates were tabulated in tables with $10^{-7} \leq n_0/n_e \leq 10^{-1}$ and $10^6 \leq n_e\tau_{recy} \leq 10^{15}$ s cm$^{-3}$.

## III. RESULTS:

We have calculated the quantities

$$\tilde{q} = \frac{Q_\|(GWm^{-2})}{n_s(10^{20}m^{-3})}\sqrt{\frac{Z_{eff}}{f_Z(\%)}\frac{\ln\Lambda}{12}} \quad and \quad \xi = n_s(10^{20}m^{-3}) \, x(100m) \sqrt{Z_{eff}f_Z(\%)\frac{\ln\Lambda}{12}} \qquad (9)$$

for $10^{-7} \leq n_0/n_e \leq 10^{-1}$ and $10^6 \leq n_e\tau_{recy} \leq 10^{15}$ s cm$^{-3}$ for Beryllium, Carbon, Neon and Argon(Figures 1—8). As an illustration of the utility of this analysis, we have estimated the requirements for detachment in high power DIII-D (20 MW) and ITER (300 MW) discharges. First, we estimate the electron temperature at the mid-plane(Eq. 7). Since $T_{sep} \sim (q_\|L)^{2/7}$, $T_{sep}$ is not very sensitive to the effects of radiation between the midplane and the divertor or to modifications in L.

$$q_\| = -\kappa_o T_e^{2.5} \frac{\partial T_e}{\partial x} = -\frac{2}{7}\kappa_o \frac{\partial T_e^{3.5}}{\partial x} : T_{es}^{3.5} - T_{div}^{3.5} = \frac{1}{\kappa_o}\int_{div}^{sep} q_\| dx \approx \frac{7}{2} L \frac{q_\|}{\kappa_o} \text{ with } L \approx 1.4\pi q_\psi R$$

$$T_{es} \approx 82.2 \, eV \left(q_\|(GW/m^2) \, R(m) \, q \, Z_{eff}\right)^{2/7} : where \, \kappa_o \approx \frac{3.1\times 10^9}{Z_{eff} \ln \Lambda}\left(\frac{erg}{cm \, s \, eV^{3.5}}\right) \quad (10)$$

For DIII-D and ITER, the predicted maximum electron temperatures (for $q_\psi=3$, $\delta \sim 1$ cm) are 120 eV and 260 eV (Table 1). One can match the appropriate $\tilde{q} = \dfrac{q(GWm^{-2})}{n_s(10^{20}m^{-2})\sqrt{\dfrac{12 f_z}{Z_{eff} \ln \Lambda}}}$ with the upstream temperature to determine the $n_o/n_e$ and/or $n_e\tau_{recy}$ required to detach the plasma.

Table 1 Typical edge temperatures, connection lengths, and parallel heat fluxes

|  | $P_\alpha$(MW) | $Q_\|$(GW/m²) | $A_\perp$(m²) | L (m) | R (m) | $T_s$(eV) |
|---|---|---|---|---|---|---|
| DIII-D 20 MW | 20 | 0.47 | 0.043 | 22 | 1.67 | 120 |
| ITER 1.5 GW | 240 | 1.5 | 0.16 | 100 | 8.00 | 260 |

Table 2 lists the results of a sample analysis of the data in the figures 3-12 for the two sample DIII-D and ITER cases. Calculations with the DEGAS code and other codes indicate that it is difficult to obtain $n_o/n_e$ much greater than $10^{-2}$, and even that is difficult except in recombining plasmas and very near the divertor plate. Impurity recycling limits can also be estimated for a divertor plasma. A reasonable upper estimate for the flow velocity of ionized impurities is the sonic speed of the background hydrogen ions which for D/T at 10 eV is $\sim 2\times 10^6$ cm/s. For a density of $\sim 10^{14}$ cm/s and L $\sim$ 3—10 m in the divertor chamber, $n_e\tau_{recy} \sim 1.5\times 10^{10}$ to $10^{11}$ cm$^{-3}$s. Higher values of $n_o/n_e$ and lower values of $n_e\tau_{recy}$ are probably unrealistic.

The data is plotted as $\tilde{q} = \dfrac{Q_\|(GWm^{-2})}{n_s(10^{20}m^{-3})}\sqrt{\dfrac{Z_{eff}}{f_z(\%)}\dfrac{\ln \Lambda}{12}}$ as a function of $T_s$ in units of 100 eV. To evaluate the necessary $n_o/n_e$ and/or $n_e\tau_{recy}$ to radiate a given $Q_\|$, we need to select a density and a value for $\sqrt{\dfrac{f_z(\%)}{Z_{eff}}} = \sqrt{\dfrac{f_z(\%)}{1+.01f_z(\%)Z(Z-1)}}$. We selected $f_z$ as 1/3 of $f_z$(fatal) defined as the impurity fraction at which the impurity radiation at 10 keV exceeds the alpha heating power. This may be somewhat pessimistic but since $Q \sim \sqrt{f_z}$, the sensitivity to the exact value of $f_z$ is not too great. From these criteria, all four impurities will require substantial enhancements due to charge exchange recombination and recycling. All four impurities require within an order of magnitude the

same level of enhancement in $n_o/n_e$ and/or $n_e\tau_{recy}$. Neon has the lowest requirements. This is because the lower Z elements have too few bound electrons, and the n=2 energy levels are too shallow. Neon has 8 electrons in the n=2 shell and Li-like and Be-like ions have greater excitation energies than C and Be. Argon does have higher emission rates than Neon, but the allowed $f_Z$ is a factor of 4 lower.

Table 2  Comparison of Be, C, Ne, and Ar radiation efficiencies for DIII-D and ITER.

| Element | Be | C | Ne | Ar |
| --- | --- | --- | --- | --- |
| Z | 4 | 6 | 10 | 18 |
| fatal fraction(%) | 14 | 6.7 | 2.4 | 0.54 |
| $0.33 \times$ fatal $f_z$(%) | 4.7 | 2.23 | 0.8 | 0.18 |
| $\sqrt{(f_z(\%)/Z_{eff})}$ | 1.73 | 1.16 | 0.68 | 0.34 |
| $Q_{\|DIII-D}/\sqrt{(f_z(\%)/Z_{eff})}$ | 0.27 | 0.41 | 0.69 | 1.38 |
| $n_o/n_e$ | $5 \times 10^{-2}$ | $10^{-2}$ | $10^{-2}$ | $5 \times 10^{-2}$ |
| $n_e\tau_{recy}$ (s cm$^{-3}$) | $10^{10}$ | $10^{10}$ | $3 \times 10^{10}$ | $5 \times 10^9$ |
| $Q_{\|ITER}/\sqrt{(f_z(\%)/Z_{eff})}$ | 0.87 | 1.3 | 2.2 | 4.4 |
| $n_o/n_e$ | $7 \times 10^{-3}$ | $8 \times 10^{-3}$ | $10^{-3}$ | $4 \times 10^{-2}$ |
| $n_e\tau_{recy}$ (s cm$^{-3}$) | $10^{10}$ | $10^{10}$ | $4 \times 10^{10}$ | $6 \times 10^9$ |

Figures 9-12 show the combined effects of impurity recycling and charge exchange recombination. There we plotted q vs. T for selected pairs of values of $n_o/n_e$ and/or $n_e\tau_{recy}$ ranging from close to coronal equilibrium ($n_o/n_e = 10^{-7}$ and $n_e\tau_{recy.} = 10^{15}$ s cm$^{-3}$) to extreme conditions of $n_o/n_e = 10^{-1}$ and $n_e\tau_{recy.} = 10^6$ s cm$^{-3}$.

Finally we solved the equations for q and $\xi$ to eliminate $T_s$ from the problem. We show q vs. $\xi$ for $n_o/n_e = 10^{-3}$ and $n_e\tau_{recy.} = 10^{10}$ s cm$^{-3}$ for C and Ne. The results are summarized in Table 3. With 1/3 of the fatal fraction and $n_s = 10^{20}$ m$^{-3}$ at the specified neutral levels and impurity recycling, DIII-D will need a connection length of 48 and 23 m for C and Ne, respectively, and ITER will need a connection length of 240 and 70 m to detach.

Table 3  x computed for $n_o/n_e = 10^{-3}$ and $n_e\tau_{recy.} = 10^{10}$ s cm$^{-3}$ for C and Ne.

| Element | q DIII-D | q ITER | $1/\sqrt{(f_zZ_{eff})}$ (100 m) | $L=x/\sqrt{(f_zZ_{eff})}$ (m) DIII-D | $L=x/\sqrt{(f_zZ_{eff})}$ (m) ITER |
|---|---|---|---|---|---|
| C | .41 | 1.3 | 0.52 | 48 | 240 |
| Ne | .069 | 2.2 | 0.85 | 23 | 70 |

## IV. Summary

Using a set of radiation loss rates for Be, C, Ne, and Ar, we have developed a set of data which can be used to set the conditions under which the energy in a scrape-off layer can be radiated. Substantial enhancements above coronal equilibrium will be needed. The conclusion of a preliminary analysis of the data indicates that Neon is the most effective radiator among the four impurities. These analyses are primarily useful to carry out a preliminary optimization of the choice of impurity species. The final choice will have to be made on the basis of further experimental studies of these impurities, and on sophisticated computer calculations.

These analyses are based on relatively simple models. Effects which might arise from flux-limited diffusion, other loss channels such as ion conduction and convection, and hydrogen recycling have not been included.

## 4. Acknowledgments

The authors are grateful for discussions with K. Lackner, R. Hulse who supplied the ADPAK code, S. Allen, G. Janeschitz, J. S. Kim and M. Rosenbluth.

Figures:

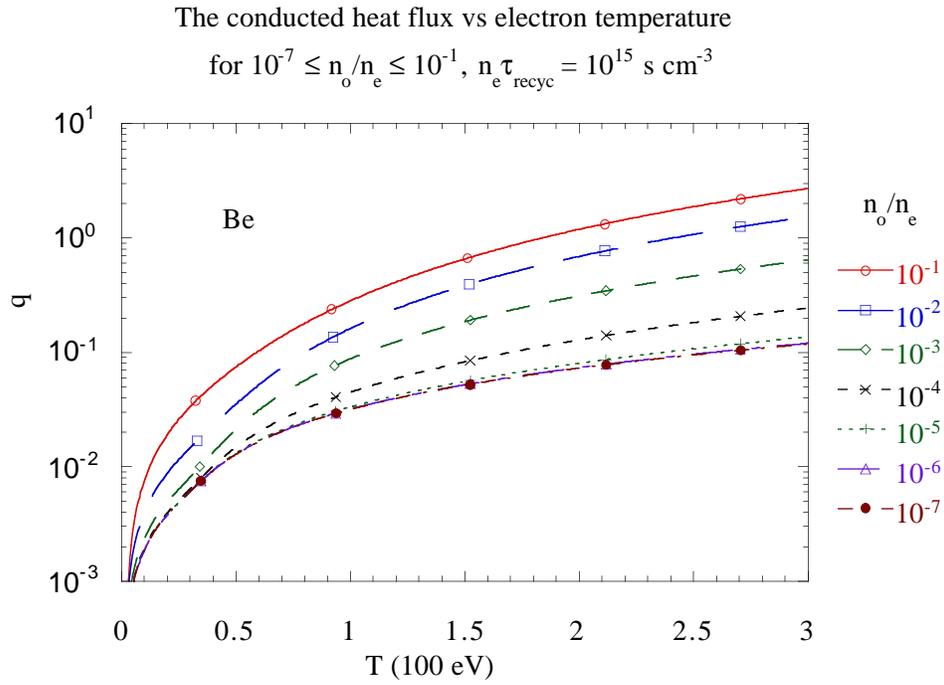

Figure 1  q vs $T_s$ for Be for $10^{-7} \leq n_o/n_e \leq 10^{-1}$.

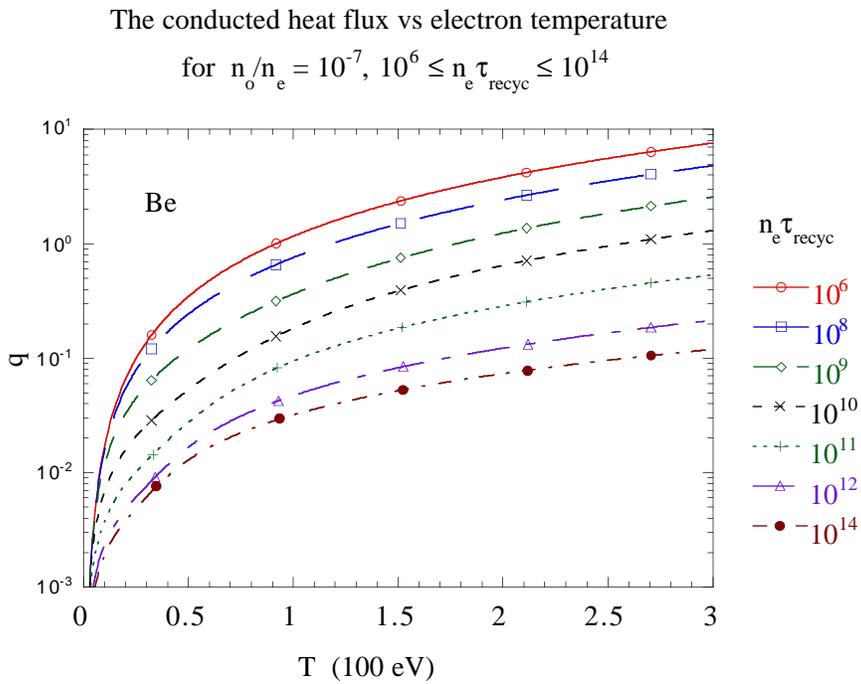

Figure 2  q vs $T_s$ for Be for $10^6 \leq n_e\tau_{recy} \leq 10^{15}$ s cm$^{-3}$.

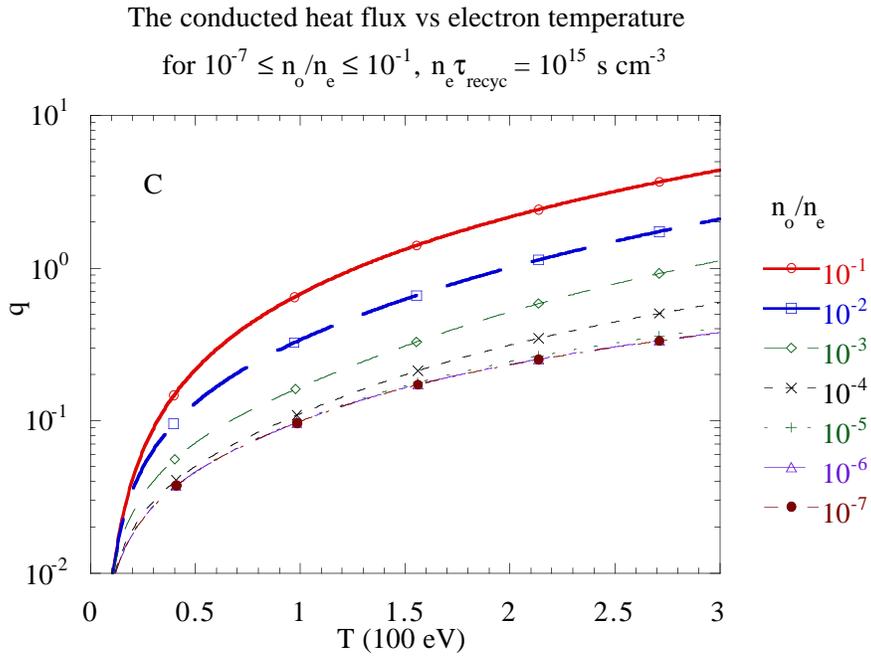

Figure 3 q vs $T_s$ for C for $10^{-7} \leq n_o/n_e \leq 10^{-1}$.

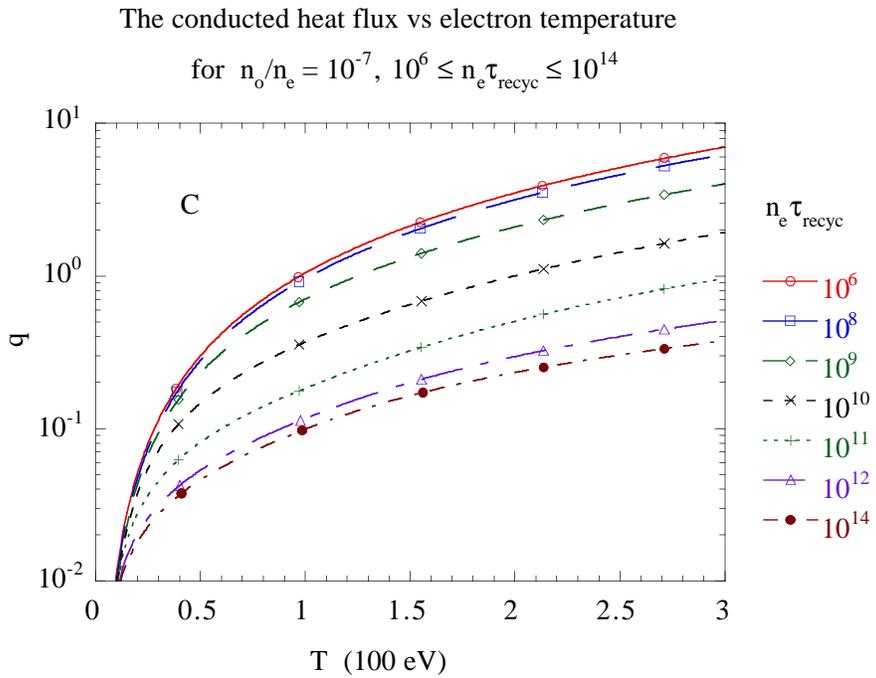

Figure 4 q vs $T_s$ for C for $10^6 \leq n_e\tau_{recy} \leq 10^{15}$ s cm$^{-3}$.

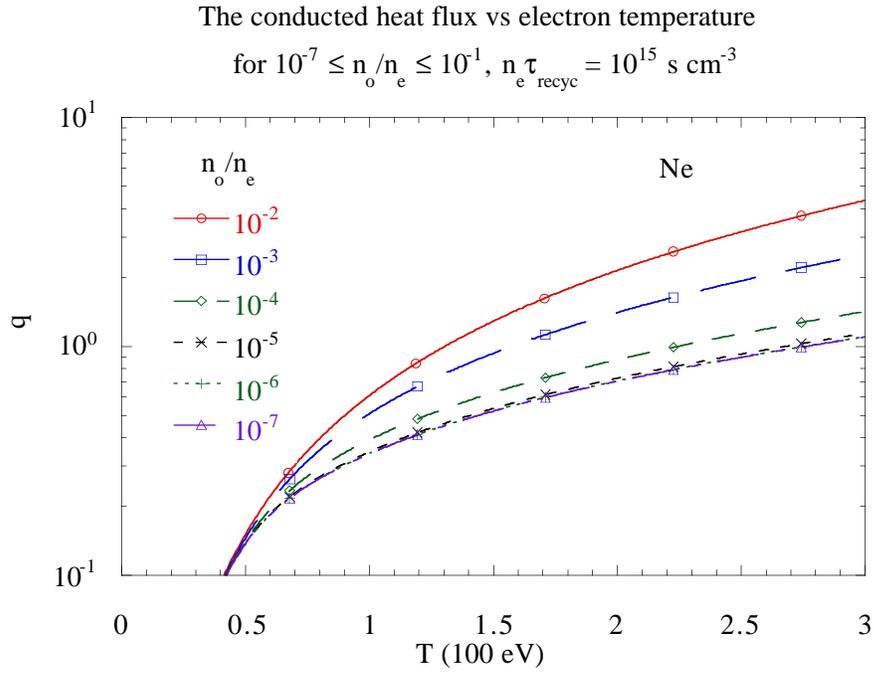

Figure 5  q vs $T_s$ for Ne for $10^{-7} \leq n_o/n_e \leq 10^{-1}$.

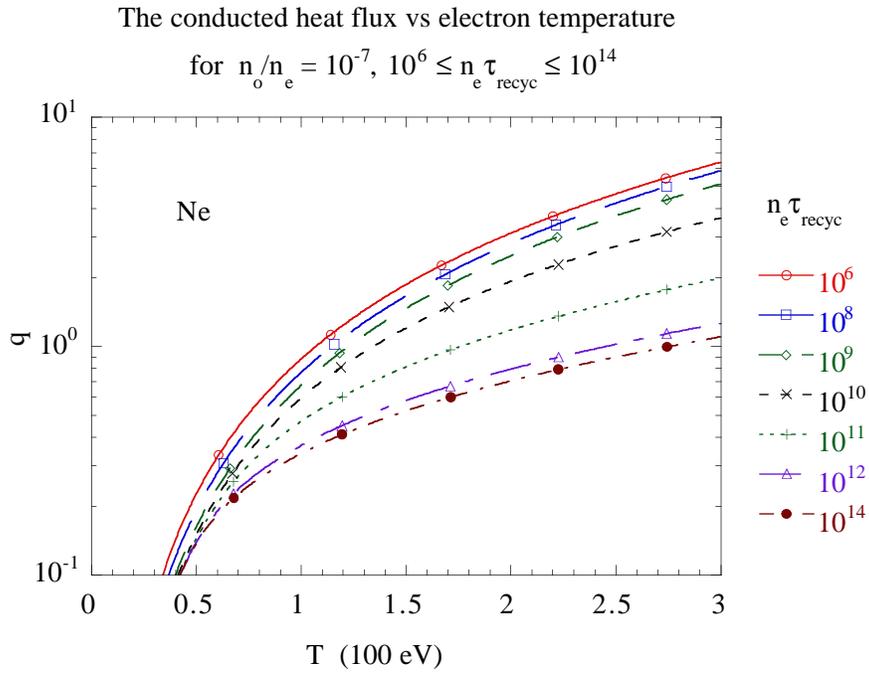

Figure 6  q vs $T_s$ for Ne for $10^6 \leq n_e\tau_{recy} \leq 10^{15}$ s cm$^{-3}$.

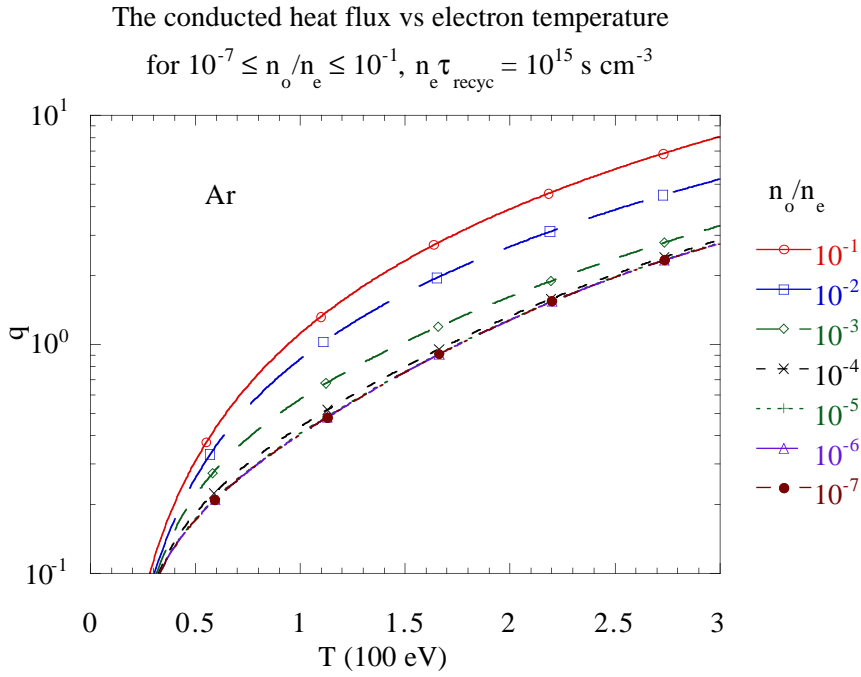

Figure 7  q vs $T_s$ for Ar for $10^{-7} \leq n_o/n_e \leq 10^{-1}$.

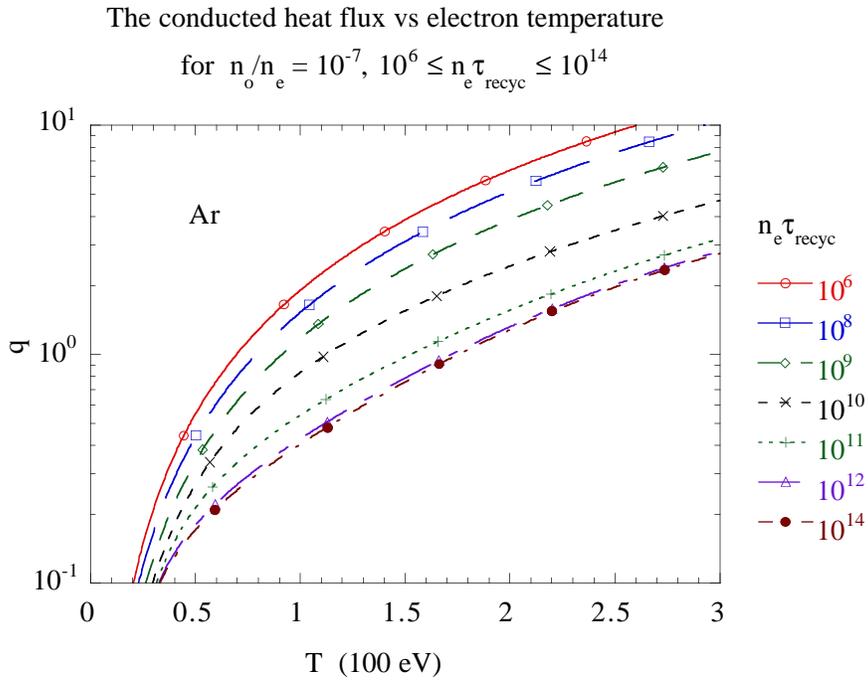

Figure 8  q vs $T_s$ for Ar for $10^6 \leq n_e\tau_{recy} \leq 10^{15}$ s cm$^{-3}$.

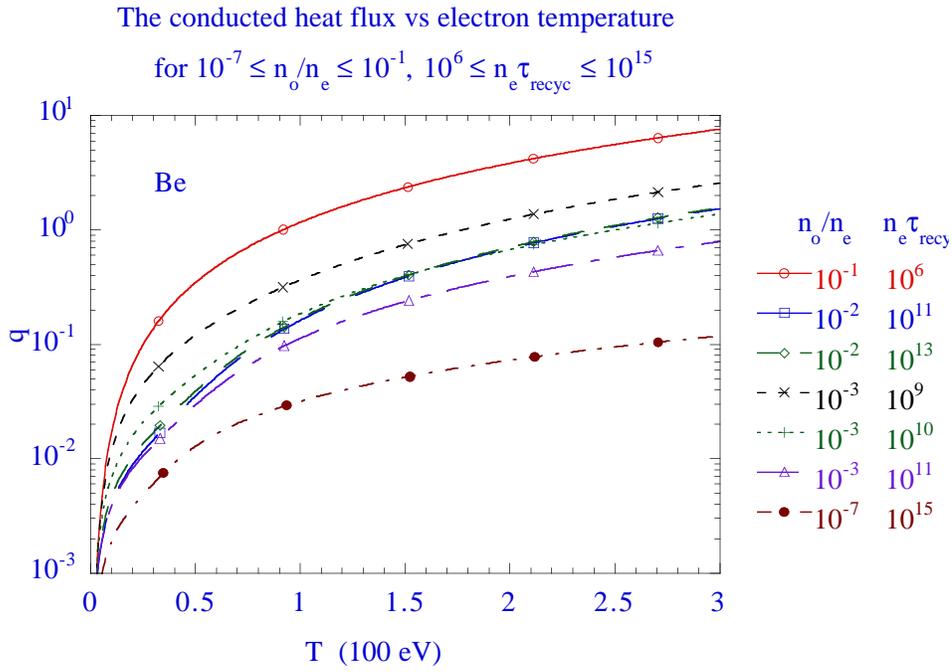

Figure 9  q vs $T_s$ for Be for selected pairs of $n_o/n_e$ and $n_e\tau_{recy}$.

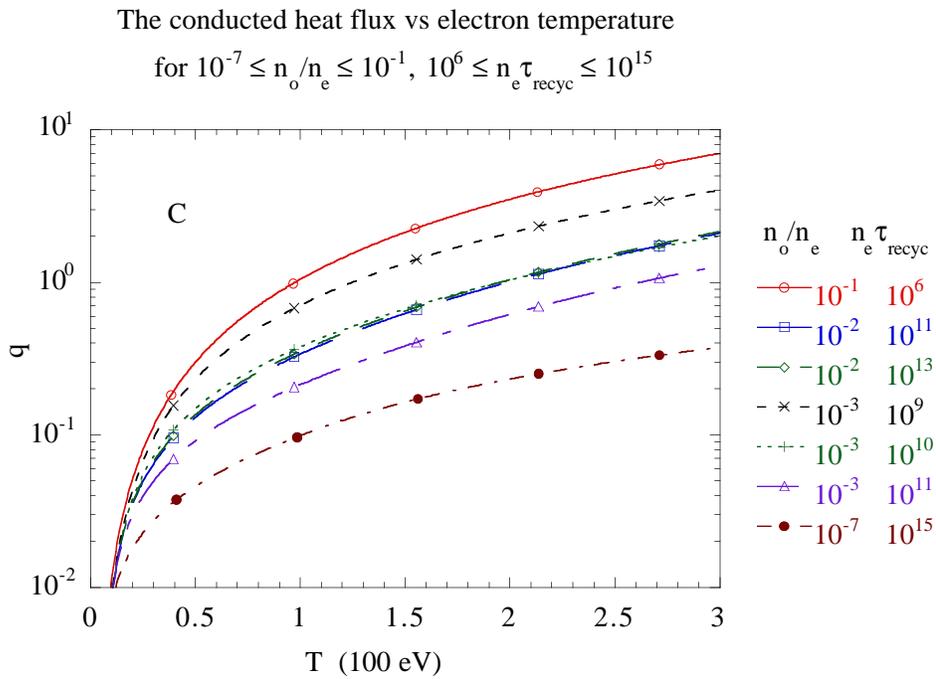

Figure 10  q vs $T_s$ for C for selected pairs of $n_o/n_e$ and $n_e\tau_{recy}$.

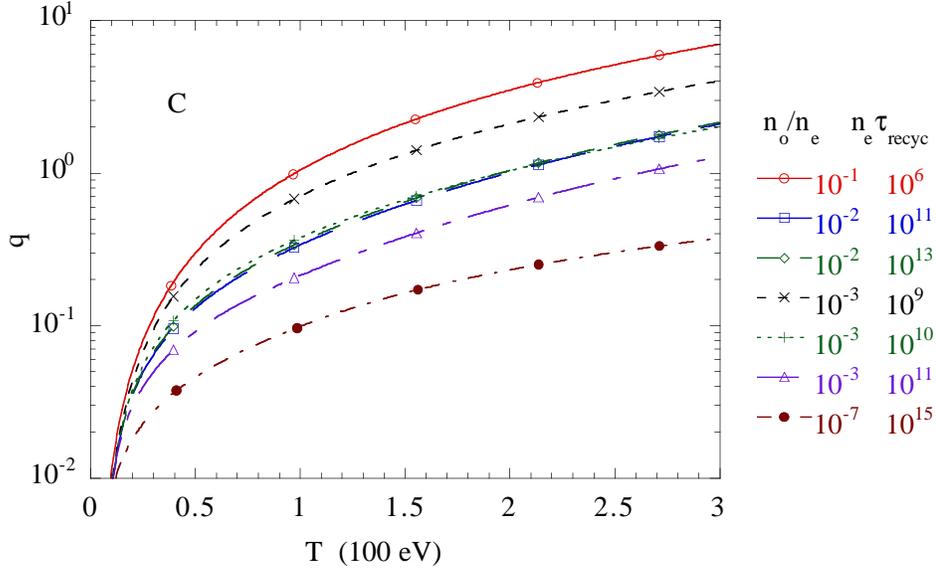

Figure 11  q vs $T_s$ for Ne for selected pairs of $n_o/n_e$ and $n_e\tau_{recy}$.

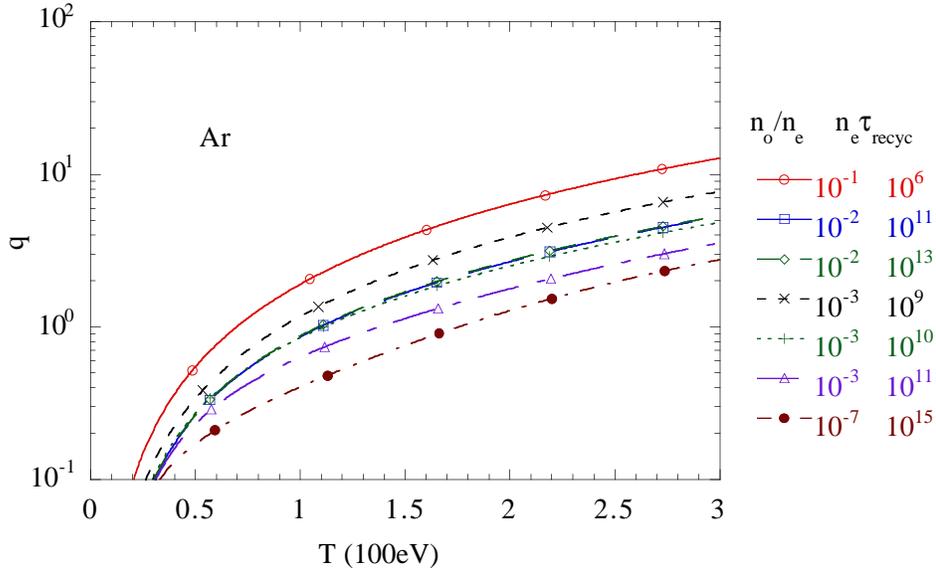

Figure 12  q vs $T_s$ for Ar for selected pairs of $n_o/n_e$ and $n_e\tau_{recy}$.

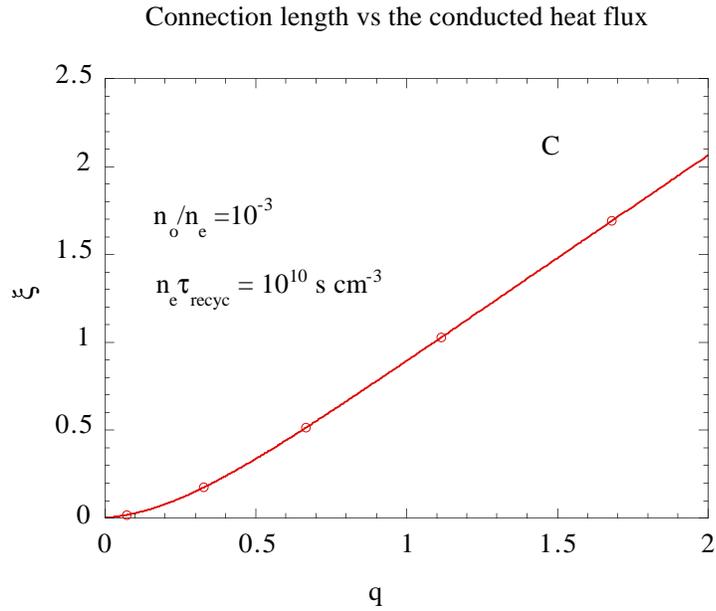

Figure 13 Normalized connection length $\xi$ vs normalized heat flux q for C for $n_o/n_e = 10^{-3}$ and $n_e\tau_{recy} = 10^{10}$ s cm$^{-3}$.

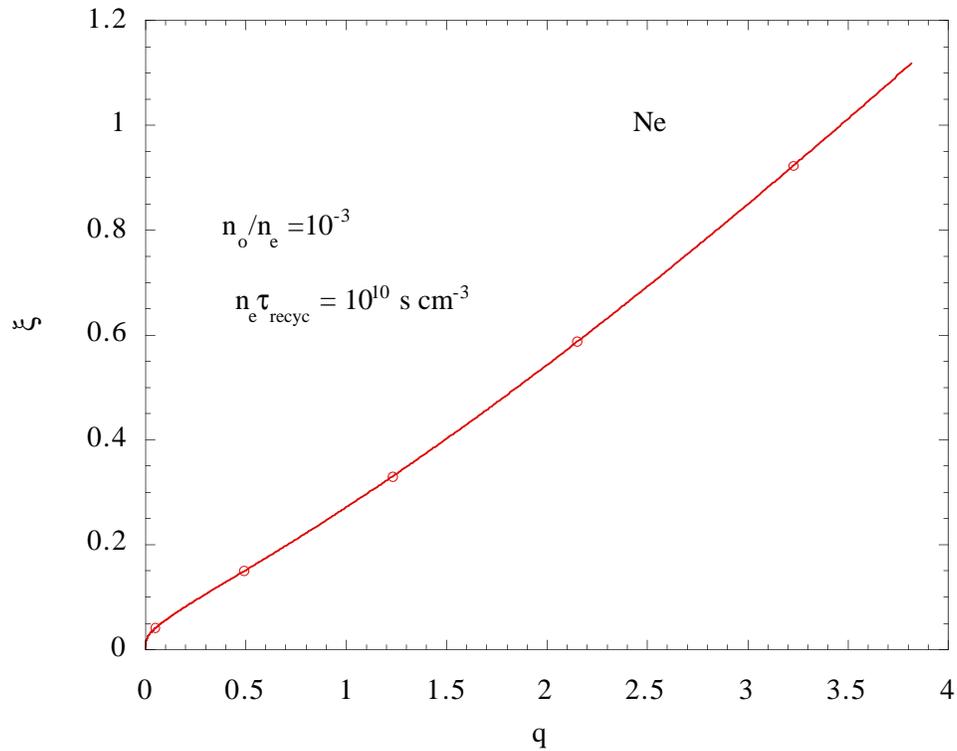

Figure 14 Normalized connection length $\xi$ vs normalized heat flux q for Ne for $n_o/n_e = 10^{-3}$ and $n_e\tau_{recy} = 10^{10}$ s cm$^{-3}$.